\documentclass[oneside,a4paper,11pt]{article}
\usepackage[margin=1in]{geometry}
\usepackage{setspace}
\usepackage{amsmath,amsfonts,amssymb}
\usepackage{algorithmic}
\usepackage{array}
\usepackage{authblk}
\usepackage[caption=false,font=normalsize,labelfont=sf,textfont=sf]{subfig}
\usepackage{textcomp}
\usepackage{stfloats}
\usepackage{url}
\usepackage{verbatim}
\usepackage{graphicx}
\usepackage{balance}
\usepackage{lineno}

\usepackage{hyperref}
\usepackage{bm}
\usepackage{cite}

\providecommand{\keywords}[1]
{
  \small	
  \textbf{\textit{Keywords---}} #1
}

\begin{document}
\title{A fully convolutional neural network to classify UXOs using multicomponent electromagnetic induction data}
\author[1]{Jorge Lopez-Alvis \thanks{now at Instituto de Geociencias, Universidad Nacional Autónoma de México}}
\author[1]{Lindsey J. Heagy \thanks{corresponding author}}
\author[1]{Douglas W. Oldenburg}
\author[2]{Stephen D. Billings}
\author[2]{Lin-Ping Song}
\author[1]{Wouter Deleersnyder}
\affil[1]{Department of Earth, Ocean and Atmospheric Sciences, University of British Columbia}
\affil[2]{Black Tusk Geophysics Inc.}


\maketitle

{\let\thefootnote\relax\footnotetext{emails: Jorge Lopez-Alvis (lopez.alvis@geociencias.unam.mx), Lindsey J. Heagy (lheagy@eoas.ubc.ca), Douglas W. Oldenburg (doug@eoas.ubc.ca), Stephen D. Billings (stephenbillings@btgeophysics.com), Lin-Ping Song (lpsong@btgeophysics.com), and Wouter Deleersnyder (wdls@eoas.ubc.ca)}


\begin{abstract}
Unexploded ordnance (UXO) clearance relies on electromagnetic induction (EMI) methods to detect and classify buried and submerged metallic objects, but the standard classification workflow requires physics-based inversion and manual expert review. We present a fully convolutional neural network (CNN) that classifies UXO directly from multicomponent time-domain EMI data, bypassing the inversion step entirely. The CNN uses a segmentation-like architecture that preserves the spatial dimensions of the input and outputs a high-resolution classification map suitable for generating a diglist. Two CNNs are used in sequence: the first detects anomalies due to metallic objects and characterizes the spatially correlated noise present in the preprocessed field data; the second performs multi-class classification using synthetic training data augmented with that noise. A physics-based parameterization of the clutter class is used to train the network to discriminate clutter from UXO across a broad range of object types. We validate this approach on two field datasets acquired with the UltraTEMA-4 system at the Sequim Bay marine test site. On a calibration grid with 21 seeded objects, the CNN detects all UXOs and discriminates two out of four clutter objects. On an independent blind grid with 35 seeded objects (17 UXOs, 18 clutter), the CNN detects and classifies all UXOs and achieves a false-positive count comparable to that of the physics-based inversion approach. We see the two methods as complementary: once trained, the CNN classifies rapidly and with minimal supervision, making it well suited as a fast first-pass tool or as an independent check on physics-based results.
\end{abstract}


\section{Background}

Clearance of unexploded ordnance (UXO) sites is important for the safety of nearby communities and infrastructure projects. Electromagnetic induction (EMI) methods are commonly used to detect and classify UXO. While the use of EMI systems on land is well established it was only recently that systems such as the UltraTEMA-4 were designed for underwater munitions \cite{funk_ultratema-4_2022}. Advanced systems aim to reduce costs related to excavation and interrogation by discriminating UXO from non-UXO objects in a single pass. These systems usually rely on many sources and receivers and create spatially dense data which have high information content since the targets are illuminated from multiple angles \cite{h_h_nelson_multisensor_2001, gasperikova_demonstration_2010, harbaugh_emi_2010, billings_ultratem_2018}.

The usual workflow for clearing a site consists of generating a map from EMI data from which anomalies of interest are picked, then classification is done for each of these anomalies and a target list is obtained \cite{beran_detecting_2013}. Picking anomalies from a gridded image created from sensor data is done by setting some threshold value of amplitude chosen to maximize the detection of ordnance expected at the site without including anomalies from sensor noise or smaller items. Once the anomalies are picked, classification is performed using a physics-based inversion approach where polarizability curves are estimated from the EMI data \cite{pasion_discrimination_2001, bell_time_2001, tarokh_statistical_2004, shubitidze_new_2008, gasperikova_uxo_2008, kappler_hybrid_2011, gasperikova_underwater_2025}. These curves are then compared with those in a library to look for a match based on some misfit measure, and a class is then assigned, see e.g.\ \cite{pasion_application_2007}.

Machine learning methods have been recently used to replace and/or complement some traditional processing in applied geophysics, see e.g.\ \cite{kim_geophysical_2018, yu2021deep}. For example, \cite{moghadas2020one} trained a CNN on a large synthetic 1D EMI dataset to estimate subsurface conductivity profiles, and \cite{wan2021robust} has examined 33 classification strategies and 11 dimensionality reduction methods to classify underground metal targets from TEM. In UXO detection and classification, neural networks have been applied for magnetic \cite{cardenas_magnetic_2022, sciegienka2025development} and acoustic \cite{williams_david_use_2021} methods with relative success. For frequency-domain EMI data, LSTM-based approaches have been applied directly to EMI signals, but they did not consider multicomponent systems and were limited to detection and clutter discrimination, no UXO-type classification \cite{lu2025unexploded}. Regarding the UXO classification with multi-component EM data, there have been a few studies which integrate neural networks in their workflows. An early study by \cite{szidarovszky_identification_2008} proposed the use of a fully connected neural network to classify UXO but its input were the inverted parameters of their polarizability curves, making this a hybrid approach. \cite{friedel_hybrid_2012} also used a neural network (self-organizing map) to classify UXOs, however, their best classification performance was achieved only when combined with a physics-based approach. More recently, \cite{heagy_machine_2020} proposed a convolutional neural network (CNN) that successfully classified a set of ordnance objects from synthetic EM data in multiple acquisition lines. The workflow we present follows this CNN-based direction; our main contributions, described below, concern the network architecture, the handling of field noise, and the treatment of clutter.

The main contributions of this work are threefold. First, and most importantly, we propose a CNN architecture suited to the high-dimensional data produced by modern multi-component EMI systems. These data carry two spatial dimensions, a temporal dimension, and multiple transmitter-receiver channels, and arranging them into a form suitable for a CNN admits many possible choices. Identifying an arrangement that respects the physics of the measurement and the geometry of the system is the main design challenge, particularly for diffusive EM data that lack the image-like structure of wave-based methods. The architecture we develop preserves the spatial dimensions of the data so that classification produces a high-resolution map, and it is flexible enough to be adapted to other multi-component or multi-channel systems. Second, we introduce a two-stage workflow in which a first CNN detects anomalies and thereby allows the spatially correlated noise present in field data to be characterized and incorporated into the training set of a second classification CNN. Third, we develop a physics-based parameterization of the clutter class that allows the network to be trained to discriminate clutter from UXO. This is crucial for machine learning methods, which require many examples of both UXO and clutter, to avoid extrapolating unreliably (classifying something not seen before). Together, these elements yield an end-to-end approach that classifies UXO directly from EMI data, which we validate using field data acquired in a marine setting.

Beyond these methodological extensions, the broader objective of this study is to investigate the potential of CNNs for UXO classification and evaluate their role alongside established classification methods. Physics-based inversion is well established and widely used within the UXO community, whereas CNN-based approaches offer distinct advantages in computational efficiency and automation. Once trained, a CNN can perform classification rapidly with minimal human intervention, making it suitable for rapid first-pass assessment of survey data. In addition, CNN-based classification can provide an independent line of evidence for evaluating classification results. Agreement between the CNN and physics-based workflows may increase confidence in the classification, whereas disagreement can identify candidates that warrant further investigation. Given the rapid development of deep learning and its increasing application in applied geophysics, it is timely to assess the potential of CNN-based approaches to complement existing physics-based workflows.

In this study, a workflow that takes EMI data and produces a classification map using two CNNs is developed; the final output is the probability of belonging to a class which is then used to select the likeliest class label and assign it to each anomaly detected. Our workflow is general enough to be applied to both terrestrial and marine data. Marine data should be pre-processed to remove the EM response of the seawater and sediments \cite{song_transient_2016,song_advanced_2020} and our approach is able to account for the remaining spatially correlated noise.

\section{Materials and Methods}

\subsection{Overview}

We propose a workflow that includes two stages: (1) detection (i.e.\ binary classification) and (2) multi-class classification. The first stage is mainly used to obtain systematic noise from field data in order to train the multi-class classifier (Figure \ref{Fig:workflow}). The output of the second stage is a classification map where the detected objects are classified either as clutter or as a UXO of a certain type (e.g. 40, 60, 81, 105 or 155 mm). In initial trials, we also considered training a classifier to distinguish UXO vs. clutter as single classes, but we observed better performance when using granular classes for the specific UXO types. We refer to this step as a multi-class classification in this work. This map may be further processed in order to obtain a diglist. The main reason for developing this two-step workflow was that the addition of systematic noise to train the multi-class CNN was crucial for obtaining accurate classification results. The CNNs are trained with synthetic data (as will be explained in the following sections) and field data are only used for validation and to approximate the systematic noise. This allows us to have a sufficiently large set of examples to train the CNNs.

\begin{figure}
	\centering
	\includegraphics[width=0.5\textwidth]{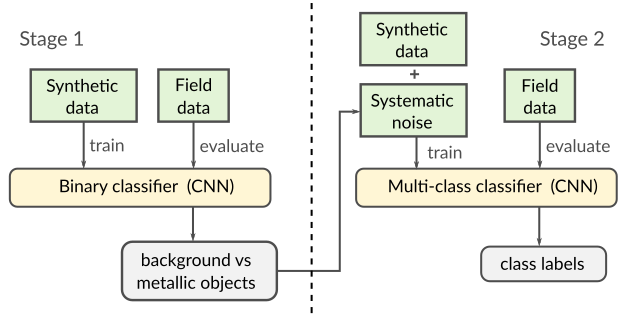}
	\caption{Our proposed two-stage workflow to detect and classify UXOs using two CNNs}
	\label{Fig:workflow}
\end{figure}

\subsection{Fully-convolutional neural network}

Neural networks are a subset of machine learning methods used to approximate nonlinear functions. The term ``deep neural network'' means that the network involves several layers that include either convolutions or dense affine operators followed by a nonlinear activation function. This general type of ``deep" network has been shown to allow for complex mappings to be accurately approximated by building a succession of intermediate (simpler) representations or concepts \cite{goodfellow_deep_2016}. Formally, deep neural networks (DNN) are mappings defined by a composition of a set of functions $\bm{\phi}_k$ as:

\begin{equation}
\mathbf{g}(\mathbf{x}) = (\bm{\phi}_L \circ \dots \circ \bm{\phi}_2 \circ \bm{\phi}_1)(\mathbf{x})
\end{equation}

\noindent where $\mathbf{x}$ is a multidimensional (tensor) input, $k = \{1, \dots, L\}$ denotes the function (layer) index, and composition follows the order from right to left. Furthermore, each $\bm{\phi}_k$ is defined as:

\begin{equation}
\bm{\phi}_k(\bm{\xi})=\psi_k(\mathbf{A}_k\bm{\xi}+\mathbf{b}_k)
\label{Eq:cnnlayer}
\end{equation}

\noindent in which $\psi_k$ is a (nonlinear) activation function, $\mathbf{A}_k$ is a matrix of weights, $\mathbf{b}_k$ is a vector of biases and $\bm{\xi}$ denotes the output of the previous function (layer) $\bm{\phi}_{k-1}$ for $k>1$ or the initial input $\mathbf{x}$ for $k = 1$. Then, training the DNN involves estimating the values for all the parameters $\bm{\theta}=\{\mathbf{A}_k,\mathbf{b}_k \mid 1 \leq k \leq K\}$ where each $\mathbf{A}_k$ or $\mathbf{b}_k$ may be of different dimensionality depending on the layer. In practice, the number of parameters $\bm{\theta}$ for such models may reach the order of $10^6$, therefore training is achieved by relying on automatic differentiation, see e.g.\ \cite{paszke_automatic_2017} and fast optimization techniques based on stochastic gradient descent (SGD), see e.g.\ \cite{kingma_adam_2017}, both usually implemented for and run in highly parallel (GPU) computing architectures.

\subsubsection{Architecture of the proposed neural networks}

In this work, we propose a segmentation architecture for the neural networks used for classification. CNNs used for segmentation are often fully convolutional which means they only include convolutional layers i.e. they do not have fully connected layers in their architecture, see e.g.\ \cite{long_fully_2015}. Having a fully convolutional CNN means that spatial information is preserved throughout the entire mapping. In our case, this means that both along-track and cross-track directions are relevant during classification e.g. a data point is more likely to be classified similar to some of its neighbors (considering these two directions). In comparison to previous approaches where the cross-track dimension was ignored in the architecture \cite{heagy_machine_2020}, our CNNs generate classification maps with higher resolution which allows us to more easily detect small objects and provides a more accurate separation and classification of objects that are close to each other.

The proposed architecture is flexible and may be adjusted to any multi-component EM system but it is easier to understand by taking the UltraTEMA-4 system as an example. The UltraTEMA-4 system has 12 receivers each with 3 components, 4 transmitters and acquires 27 time channels. Figure \ref{Fig:ultratemacycling} shows the UltraTEMA-4 system as deployed in the field and the physical dimensions of its layout.

\begin{figure}
	\centering
	\includegraphics{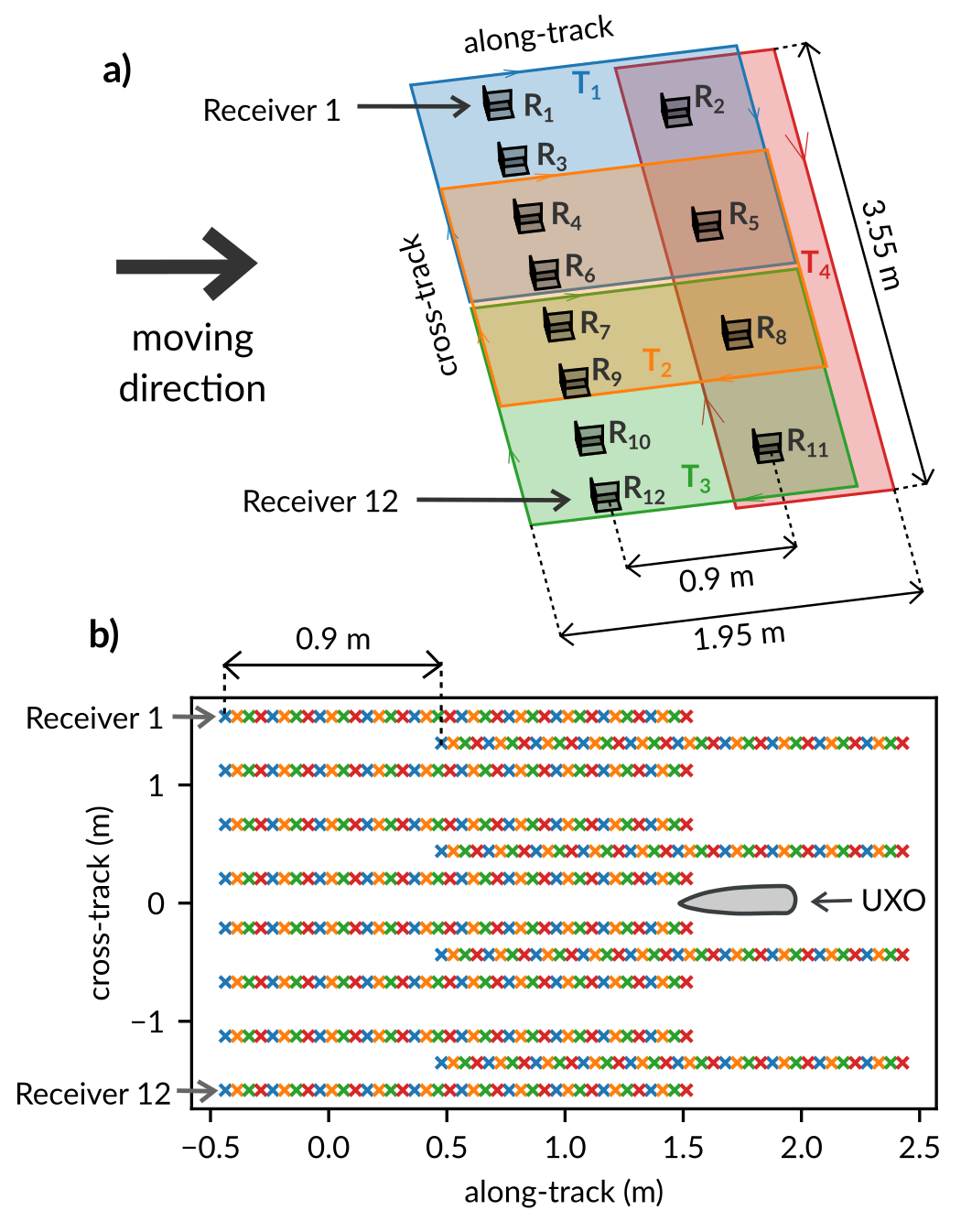}
	\caption{(a) Layout of the UltraTEMA-4 system and its physical dimensions (b) sketch of the acquisition points showing how transmitters are cycled (each color denotes the firing of a particular transmitter). Receivers 2, 5, 8 and 11 lead the others by about 0.9m, which causes the spatial offset seen in the bottom figure.}
	\label{Fig:ultratemacycling}
\end{figure}

For image datasets, a common application of segmentation CNNs involves assigning a label to each pixel of an image, as opposed to traditional classification where we get a single label per image. Such images may be composed of several channels, for example, an RGB image, where each of the channels is meant to convey independent information. We aim to analogously create an “image” from the EM data for which segmentation would then entail labeling each “pixel”. Defining the input of our CNN (that is, creating such an “image”) requires some design choices on how to order the EM data. First, following a similar approach to \cite{heagy_machine_2020}, we choose to process the EM data using a fixed spatial window in the along-track dimension, this means that we are considering a fixed number of transmitter cycles to create our input. Our proposed architecture then preserves the two spatial dimensions (cross-track, nrx, and along-track, nx, directions) as well as the temporal dimension (nt). This defines a three-dimensional “image” with dimensions: nx x nrx x nt. We apply the same normalization and data scaling steps described in \cite{heagy_machine_2020} where we scale the data as a function of time to account for the exponential decay of EM signals, and second, we scale by the maximum amplitude of each sample so that the data all lie within [-1, 1]. Information regarding different receiver components and different transmitters define the channels of the input (analogous to the RGB channels of an image; they are labeled as “ntx x 3” in Figure \ref{Fig:architecture}). This means that our input is three-dimensional with multiple channels. However, to create a classification map we are only interested in preserving spatial dimensions, so within the architecture, we eventually collapse (or reshape) the temporal dimension and end up with a 2D output. Note that for the output, the number of channels is now the number of classes we want to distinguish. Figure \ref{Fig:architecture} shows our proposed architecture: the input and output have the same width and height, that is the number of along-line positions (nx) by the number of receivers (nrx). Here, we can also see that only convolutional layers are used which makes our architecture fully convolutional. Note also that there are two convolutional stages: an initial one where 3D convolutions are applied, and one after the “reshape” operation where 2D convolutions are used.

\begin{figure*}
	\centering
	\includegraphics[width=\textwidth]{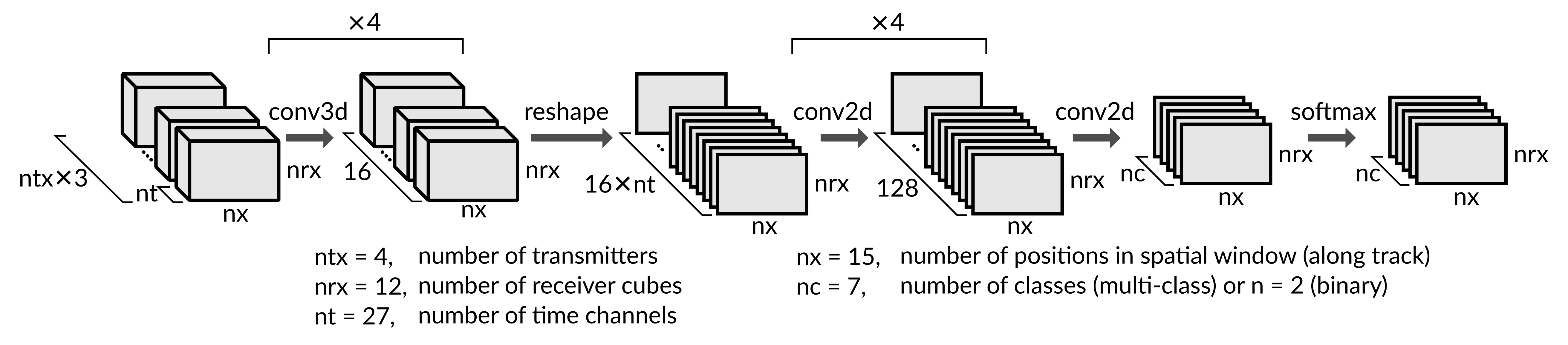}
	\caption{Architecture of CNNs for classification and detection. The first part (to the left of “reshape”) indicates input and intermediate representations are 3D while after “reshape” intermediate representations and final output are 2D. The number of hidden features is 16 for the first part and 128 for the second part. The heading “x 4” indicates that the layer repeats four times (with the same dimensions).}
	\label{Fig:architecture}
\end{figure*}

The activation functions $\psi_k$ involved in all layers (except the last layer where softmax is applied directly) are rectified linear units (or ReLU) and are applied element-wise (or pixel-wise):

\begin{equation}
 \psi(x) = \begin{cases} x & \text{if } \; x>0, \\ 0 & \text{otherwise}, \end{cases}
\end{equation}

\noindent and the output of the last convolutional layer, which has a number of channels (nc) equal to the number of classes defined, is translated to probability values using a softmax function:

\begin{equation}
 \psi(x)_i = q_i(x) = \frac{e^{x_i}}{\sum_{j=1}^{nc} e^{x_j}}
\end{equation}

\noindent where $q_i(x)$ denotes the probability of class $i$ for pixel $x$.

In this way, the output of the CNN is given in terms of a series of probability maps equal to the number of classes. Each of these maps gives the probability of a certain “pixel” to be classified as a particular type of object i.e. for a single pixel the probabilities should add up to one. To assign a class for a fixed window of data we can simply take the class with the highest probability.

The UltraTEMA-4 system is usually set up to operate at 30 or 90 Hz base-frequency to avoid 60 Hz noise from the power lines. Data collection uses stacking and that translates into one stacked measurement every 7.5 cm approximately \cite{song_electromagnetic_2024}. The system acquires data from each different transmitter sequentially (Figure \ref{Fig:ultratemacycling}), but for our CNN, we wish to use all of this information together so we use the midpoint of all 4 transmitter positions as the reference point for our classification. Thus, information associated with different transmitters is put in the channels of the input of our CNN. Since each successive transmitter produces measurements approximately every 7.5 cm, the error associated with placing our output in the middle point is considered low, but it may bias the position of detections and classifications up to 2 times this distance (around 15 cm for the field case considered). A fixed number of transmitter cycles equal to 15 was chosen for our CNN input, which approximately spans 4.5 m in the along-track direction. The number of classes for the last layer of the detection (binary) CNN is 2 while for the multi-class CNN it is set to the number of UXO classes + 2 (background and clutter). In our work, we will use 7 classes as explained in section \ref{Sec:trainingsynth}. Therefore, the different dimensions in the architecture in Figure \ref{Fig:architecture} are ntx = 4, nrx = 12 x 3 components, nt = 27, nx = 15 and nc = 2 for the first stage CNN distinguishing metallic targets from background while nc = 7 for the second stage CNN classifying targets.

Implementation and training of the proposed CNNs was done using the open-source library Pytorch \cite{paszke_pytorch_2019}. To perform the many tests mentioned in this work we used GPUs which reduced the training time of the CNNs significantly.

The architecture we have described is a key contribution of our study. While CNNs have been studied extensively on traditional image datasets, their application to multi-component and multi-channel geophysical data has received less attention. The example we provide here is for multi-component EMI data, but the same approach may be applied to other geophysical data that are collected and processed similarly, i.e. where spatial and/or temporal structure in the data can be exploited during processing. For wave-based methods (sonar, seismic, radar, etc.), the data already have an image-like structure, so casting them into a form suitable for a CNN is comparatively direct. Diffusive EM data do not share this structure, and a central part of our work is restructuring the input so that the physics of the measurement and the geometry of the system are respected. The architecture also adapts readily to other system configurations: a different number of transmitters, receivers, or receiver components changes the size of the corresponding input dimensions and channels without altering the overall structure, so the same design can accommodate other multi-component or multi-channel EMI systems.

\subsubsection{Defining segmentation-like labels for the training dataset}

As mentioned above, the CNN output has the same spatial dimensions as the input, therefore we need to define labels with these dimensions (nx by nrx) to be used as target labels during training. To convert our synthetic EM data into these target labels we propose the following approach. First, we choose a single time channel (time channel \#6, in our case, because we expect this intermediate channel to highlight all metallic targets and to have a good signal to noise ratio) and normalize the input by the maximum amplitude across all transmitter-receiver pairs. Next, we compute the magnitudes per transmitter (using the x-, y- and z-components of the receivers) for each “pixel” of the input and then add them up across transmitters. That is, we apply the following equation per data point $d$:

\begin{equation}
\sum_{i=1}^{ntx} \sqrt{d_{x,i}(t)^2 + d_{y,i}(t)^2 + d_{z,i}(t)^2}
\end{equation}

\noindent where $i$ denotes the transmitter number and $t$ denotes the time channel.  Finally, we define a threshold and ``pixel" values above this threshold are labeled according to the class used to generate the synthetic data (Figure \ref{Fig:labelmasks}); the remaining pixels are simply labeled as background pixels (blue in Figure \ref{Fig:labelmasks}). Note that for the present work, we have only considered a single object per fixed spatial window, therefore the current version of our proposed workflow is not able to account for multiple objects close to each other (within the same spatial window); we will refer to this setting as the multi-target scenario in the remainder of the text. Future research will examine multi-target scenarios.

\begin{figure}
	\centering
	\includegraphics[width=0.5\textwidth]{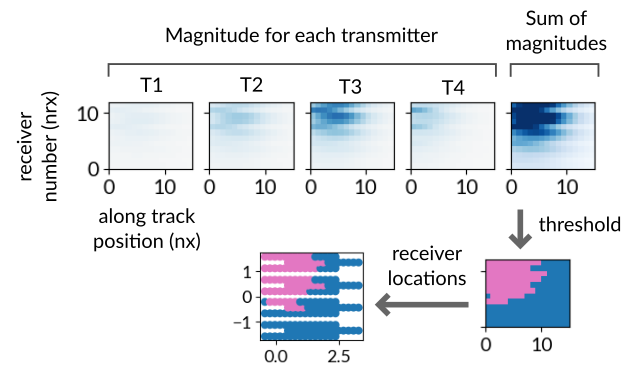}
	\caption{Procedure to assign segmentation-like labels to train our classification CNNs. In the rightmost part of the figure, points in pink denote signal from a metallic object (UXO or clutter) while points in blue denote background.}
	\label{Fig:labelmasks}
\end{figure}

\subsubsection{Processing a line of data}
\label{Sec:processingline}

According to the proposed CNN architecture, the input is constructed using a fixed window in the along-track direction. Then, the full acquisition line is processed by sliding this window with a defined step and performing classifications at each step. For our workflow, we considered a step equal to 1. In this way, the outputs of the CNN will have an overlap of $nx-1$. To generate the classification map we use a voting scheme of all overlapping outputs and assign the label that is voted the most. A similar procedure is used to average probabilities across the sliding window applying the equation to each data point (or “pixel”):

\begin{equation}
q_{avg}^i = \frac{1}{nw} \sum_{k=1}^{nw} q_k^i,
\end{equation}

\noindent where $q_{avg}^i$ is the average probability for class $i$, computed using equation 6, and $nw$ is the number of overlapping windows for that particular data point. Note that the probability outputs are only needed for post-processing, i.e., assigning a single label per detected object and/or a diglist.

The sliding window helps in two ways: (1) pixel or data points in the central part of the window are usually classified with more confidence therefore it is better to pass through all the acquisition line stepping one by one so that all data points are allowed to pass through the central part of the window; (2) the voting scheme for overlapping windows makes the approach more robust i.e. if a single classification is performed the class assigned to some points could be incorrect, voting usually fixes this because the data are processed in different along-track positions within the window. This sliding window might cause some edge artifacts (less voting and not letting some data points pass through the central part of the window) however, in practical terms, we expect the impact to be minimal because acquisition lines are usually designed to start a few meters beyond the region of interest, which is enough for our method to cover the span of a full window (~4.5 m where a typical full acquisition line length is at least ~100 m). The window size should be chosen small enough to avoid including more than one object as explained above; this constraint also helps to minimize the length of these edge effects.

\subsection{Generating the training dataset}

\subsubsection{Dipole forward model}

Since UXOs are compact objects, their secondary magnetic fields $\mathbf{B}_s$ can be described as caused by an induced magnetic dipole moment $\mathbf{m}$ \cite{pasion_discrimination_2001}. For the UltraTEMA-4 system, the UXO electromagnetic response is measured in terms of $d\mathbf{B}/dt$ which is modeled with:

\begin{equation}
\frac{d\mathbf{B}_{s}}{dt} = \frac{\mu_0}{4 \pi} \Bigg [ \frac{3 \hat r \big ( \hat r \cdot \frac{d\mathbf{m}}{dt} \big ) - \frac{d\mathbf{m}}{dt}}{r^3} \Bigg ].
\end{equation}

For an ordnance object, $d\mathbf{m}/dt$ is approximated using a polarizability tensor $\mathbf{L}(t)$. If we align our coordinate system with the principal axes of the UXO using a rotation matrix $\mathbf{A}$, the polarizability tensor $\mathbf{L}(t)$ is diagonal and the time-derivative of the induced dipole moment is obtained from:

\begin{equation}
\frac{d\mathbf{m}}{dt} = \mathbf{A^T L}(t) \mathbf{A h}_p,
\end{equation}

\noindent where $\mathbf{h}_p$ is the primary magnetic field generated by the transmitter at the location of the UXO just before the current is shut-off and the diagonal polarizability tensor is:

\begin{equation}
\mathbf{L}(t) = \begin{bmatrix} L_{1} (t) & 0 & 0 \\ 0 & L_{2} (t) & 0 \\ 0 & 0 & L_{3} (t) \end{bmatrix},
\end{equation}

Since UXOs are usually axisymmetric objects, $L_3$ is associated with a dipole parallel to the largest (longitudinal) dimension of the UXO. Eddy currents about this axis decay slower than those associated with the other two transverse axes which are usually equal i.e. $L_1=L_2$ (Grimm, 2003).

According to this dipole model, we require the polarizabilities, orientation, and location of the UXO to simulate the response. A library of polarizabilities for a large collection of ordnance objects is available from the U.S.\ Department of Defense \cite{murray_munitions_2016}. This library was queried to generate the synthetic data used to train the neural networks in the proposed workflow. To run the forward simulations, we use the simulation code that is a part of BTField software \cite{black_tusk_geophysics_btfield_2025}.

\subsubsection{Parameters for the training data}

The proposed neural networks need to be trained with all possible UXOs that may be encountered in the field site. This includes defining a range for the following parameters:

\begin{itemize}
\item type of object (UXO, clutter or no object)
\item location of the UXO below the system
\item orientation of the UXO
\item background noise
\end{itemize}

Since the neural networks process the data using a fixed spatial window, the training set needs to include examples where this spatial window includes: (1) a UXO, (2) a clutter object, and (3) no object (only the background signal). The type of UXO is taken from a set of possible ordnance objects that may be defined by means of historic records or any other field-specific prior information. In general, we do not know what clutter objects may be encountered. Therefore, a useful strategy is to define a clutter class that is as diverse as possible. The need for including diverse clutter examples during training of the classifier has been identified in previous works \cite{friedel_hybrid_2012}. To define these clutter samples we developed a strategy detailed in the following paragraphs.

Our clutter design strategy uses a physics-based parameterization to define 9 parameters per clutter object. This parameterization models the EM decay in time for three different orthogonal directions aligned with the geometry of the object, i.e., $L_1$, $L_2$ and $L_3$ are each modeled with the following equation:

\begin{equation}
L(t) = kt^{-\beta} \exp{(-t/\gamma)},
\label{Eq:pasion}
\end{equation}

\noindent where $k$, $\beta$ and $\gamma$ control the behavior of the EM decay associated with one of the axes \cite{pasion_discrimination_2001}. For each UXO type, we extract the polarizability curves from the library and carry out a nonlinear parametric inversion to find the 9 parameters that best fit each curve. Since the parameters $k$ span several orders of magnitude, they are first log-transformed. Then, in the 9-dimensional parameter space, we create some random samples from Gaussian distributions whose means are given by the fitted parameters and their standard deviations are assigned a value of five percent of the range of that parameter  (Figure \ref{Fig:parspace}b shows an equivalent procedure for a two-dimensional space to make visualization easier, i.e. in this figure we assume there are only two parameters involved $\beta_1$ and $\kappa_2$). These samples may be translated to polarizability curves that are used in the dipole forward model.

Finally, we wish to use the remainder of the parameter space to build our clutter class. To do this, we define a distance threshold to ensure that clutter objects are sufficiently far from UXO in the nine-dimensional parameter space defined by $p$. The distance in parameter space is computed after dividing each parameter by the standard deviations previously defined; this weighted distance is used to account for the differing scales of the parameters i.e. if a parameter has a wider range, the distance along that axis should be longer. The minimum and maximum values for the ranges in the parameter space were taken by fitting this model to objects in the library that were smaller (20mm) and larger (100-lb bomb) than those targeted to be classified. To define the distance threshold we randomly sample the parameter space within these ranges and then compute the minimum distance from any UXO for each of these samples:

\begin{equation}
\min_{p_i} \left(\sqrt{(\mathbf{p} - \mathbf{p}_i)^2}\right), \quad \mathbf{p}_i \in P_{UXO},
\end{equation}

\noindent where $\mathbf{p}$ denotes a sample in the parameter space (normalized by the standard deviations) and $\mathbf{p}_i$ are the samples from the UXO set $P_{UXO}$. Then, using a histogram of such distances we choose a threshold value that discards the closest samples (Figure \ref{Fig:parspace}a). An example of the polarizability curves obtained for a single clutter object is shown in Figure \ref{Fig:clutterpolcurves}. Notice that the gray lines (clutter object) are far from any combination of colored lines (UXO objects) as expected.

\begin{figure}
	\centering
	\includegraphics[width=0.5\textwidth]{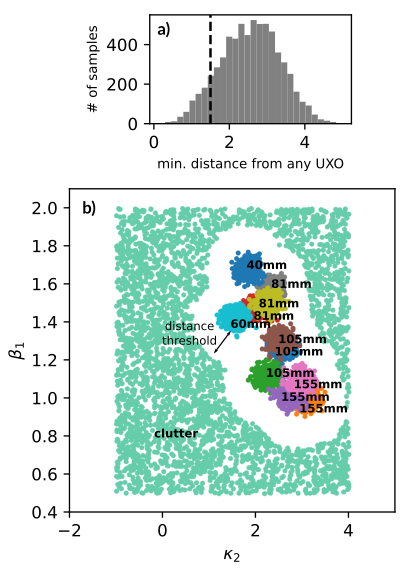}
	\caption{(a) Distribution of minimum distances w.r.t. to UXO objects in the parameter space and selected distance threshold (dashed line). (b) Two-dimensional sketch to illustrate the process to define clutter objects in the 9-dimensional parameter space.}
	\label{Fig:parspace}
\end{figure}

\begin{figure}
	\centering
	\includegraphics[width=0.5\textwidth]{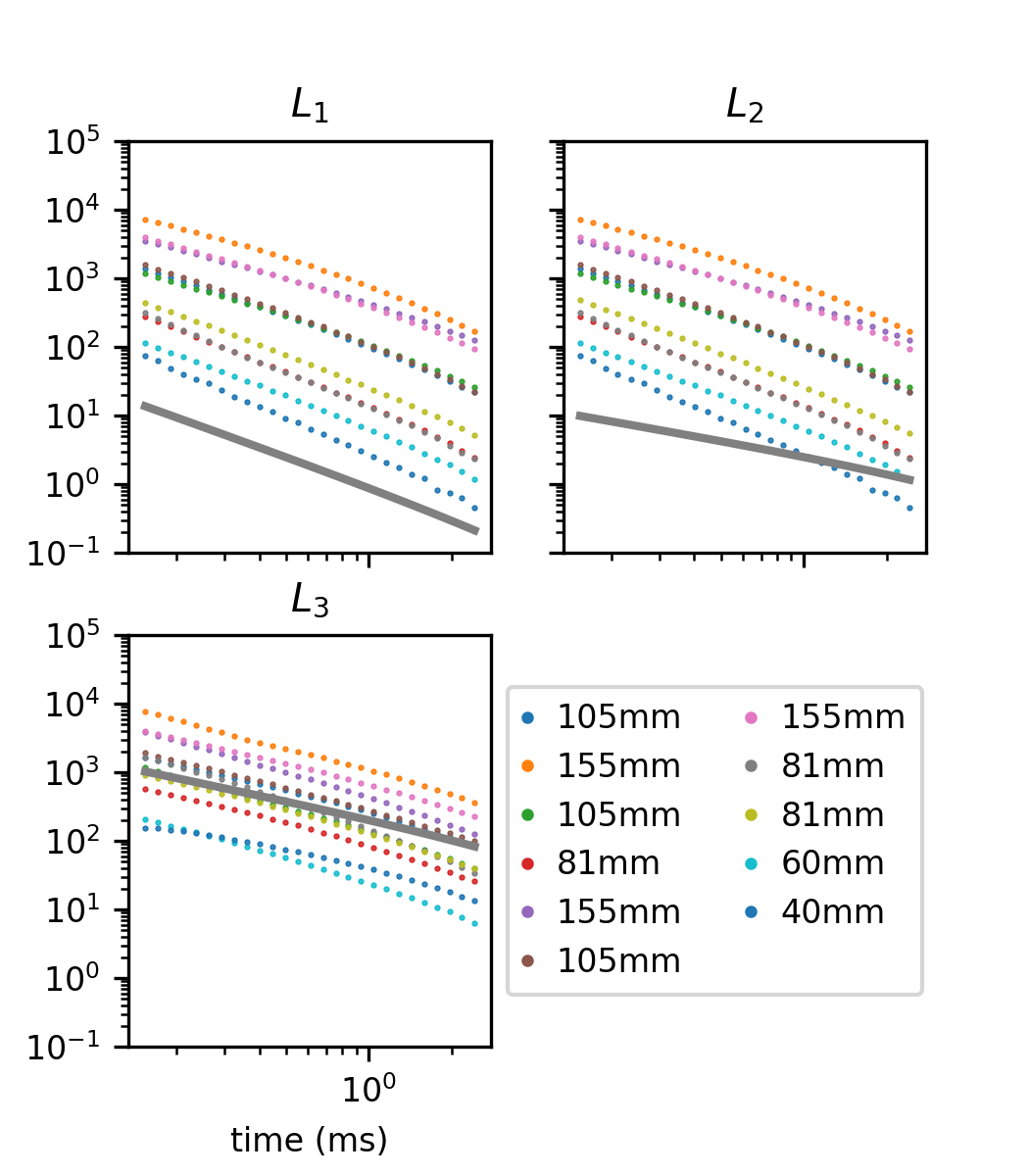}
	\caption{An example of the polarizability curves for one sample of a clutter object (gray continuous lines) generated according to the physics-based strategy. Dotted lines in color are the polarizability curves from selected UXO objects.}
	\label{Fig:clutterpolcurves}
\end{figure}

We tested our approach using a synthetic example. We simulate a multi-line acquisition and include both UXO and clutter objects. Some manually designed clutter objects were used to test the robustness of the clutter design strategy in identifying clutter. These manually designed clutter were built by taking polarizability curves from the library but combining $L_1$, $L_2$ and $L_3$ so that they avoid typical dimensions of UXO objects e.g. we considered an object which had $L_3$ of a 155mm but the $L_1$ and $L_2$ of a 40mm in order to approximate a long rod-like metallic object.  The CNNs correctly identified all clutter objects in this test.

When generating the training dataset, we need to assign the location of the object (UXO or clutter). To do this, we define a rectangular region based on the length of the spatial window in the along-track dimension and the width of the system; this is shown by the colored region in Figure \ref{Fig:labelingbox}a and by the solid black line in Figure \ref{Fig:labelingbox}b. The height of the system above the sea-bottom is defined as a survey parameter and we use this to define a minimum vertical offset between the system and the shallowest target. The depth of the region is chosen according to the expected depths of the UXO/clutter objects in the field (a maximum of 0.5 m). We randomly choose a location in the region to simulate data associated with UXO or clutter objects (e.g., pink dots in Figure \ref{Fig:labelingbox}b). Objects outside of this window can still generate measurable signal and in early tests, we noticed that this can lead to ``halo" effects during classification: acquisition lines that passed close to objects, but not directly above them, classified pixels in similar along-track distances either as a UXO or clutter object (not necessarily the same type as the line that passed directly above the object). Since we want our classification map to pinpoint the location of the UXOs such “halo” effects are undesirable and we aim to reduce them. In order to reduce these effects, we design additional training samples where UXO or clutter objects are further away from the system and all points should still be labeled as background. These points are shown in red in Figure \ref{Fig:labelingbox}b. By including these additional samples that are labeled as background, we found that the "halo" effects were reduced and our classification map better localizes the objects. A sampling of objects and the associated classification are shown in Figure \ref{Fig:labelingexamples}; note that when objects are too far from the center all the points are labeled as background regardless of the type of object. However, the 105 and 155 mm ordnances which are slightly outside of the window, are still correctly classified i.e. the classifier will produce a good result even though only a portion of the EM signal is available. The orientation for UXO and clutter objects was chosen to span from 0 to 2$\pi$ for yaw, pitch and roll. Figure \ref{Fig:synthdata} shows the synthetic EM data for a single training sample.

\begin{figure}
	\centering
	\includegraphics[width=0.5\textwidth]{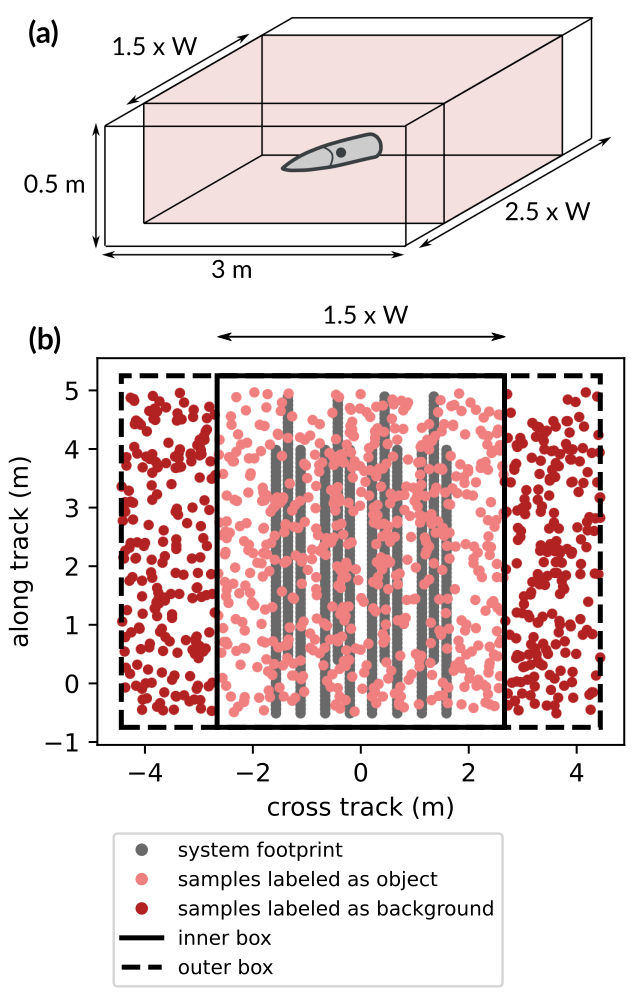}
	\caption{(a) Box to randomly place the object (UXO or clutter) while generating the synthetic training dataset, where W denotes the width of the system (see Figure \ref{Fig:ultratemacycling}). (b) 1000 random samples of location for objects (red and pink dots) and the corresponding system footprint in the defined spatial window in the along-track dimension (grey dots).}
	\label{Fig:labelingbox}
\end{figure}

\begin{figure}
	\centering
	\includegraphics{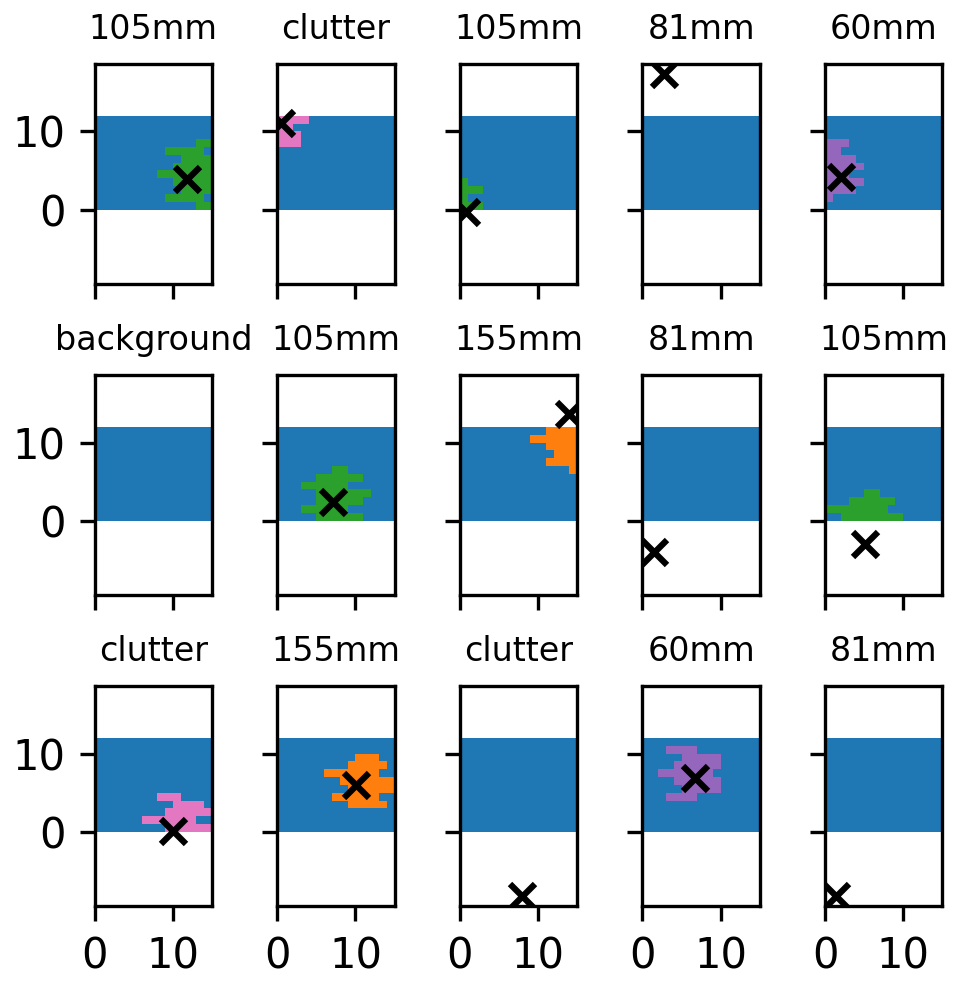}
	\caption{Examples of labeling for random synthetic samples. Axes give horizontal position (pixel number), ‘x’ gives the center of the object and the label at the top of each plot denotes the type of object that was used in the forward model. Different colors correspond to the label on top of each inset (see Fig. \ref{Fig:clutterpolcurves} for all the color labels).}
	\label{Fig:labelingexamples}
\end{figure}

\begin{figure}
	\centering
	\includegraphics[width=0.475\textwidth]{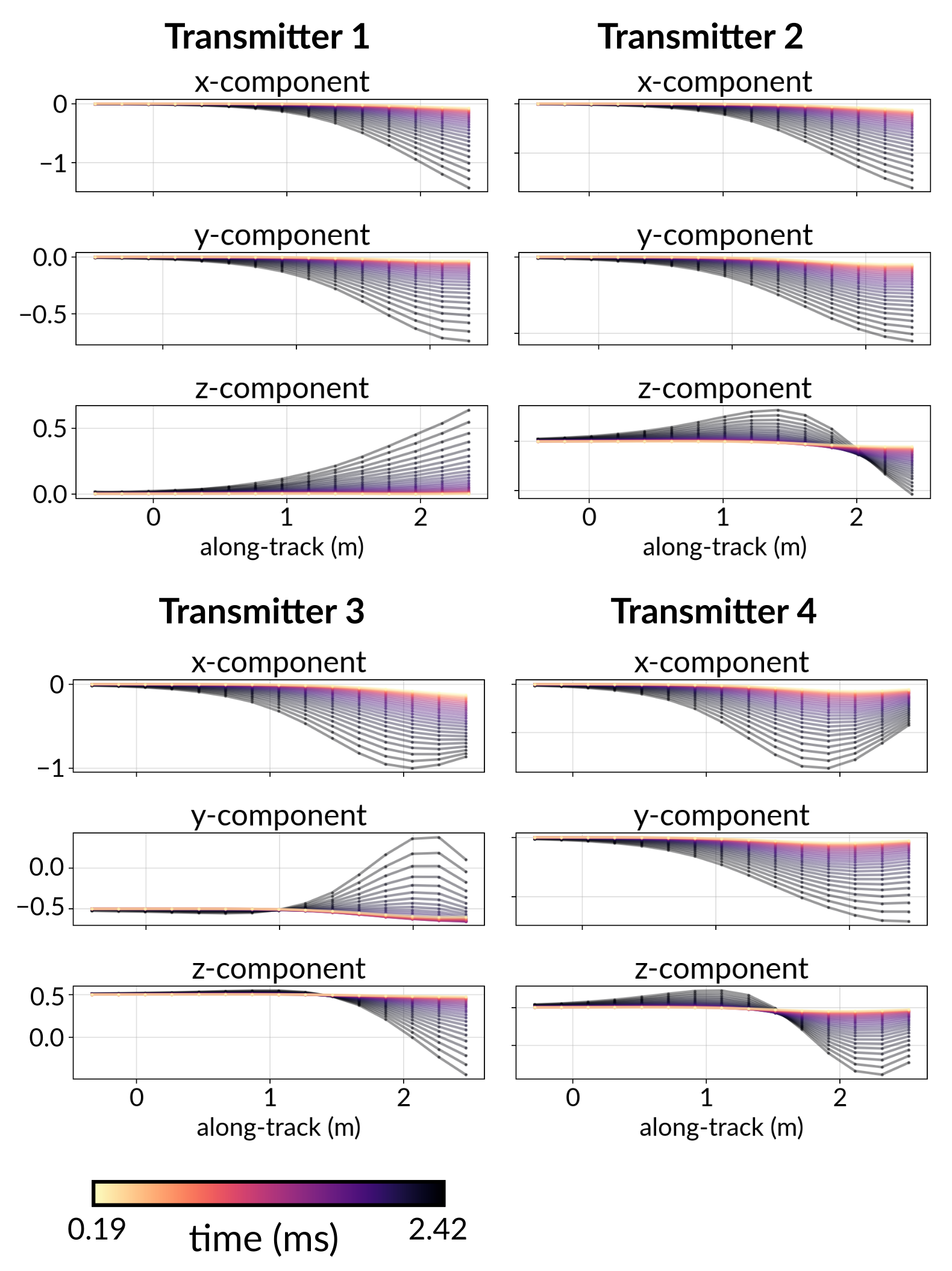}
	\caption{Synthetic EM data generated for a single training sample (the color of the text denoting each transmitter is the same as in Figure \ref{Fig:ultratemacycling}).}
	\label{Fig:synthdata}
\end{figure}

We considered data that were pre-processed to remove the EM response of the seawater and sediments \cite{song_transient_2016,song_advanced_2020}. The background noise added to the training dataset was defined differently for the neural network of the first stage (binary classification) than that for the neural network of the second stage (multi-class classification). Since our aim for the first stage is only to discriminate between background and object, we used a background noise modeled only from the standard deviation signature of the system. This signature was estimated from field data by selecting areas sufficiently far from anomalies due to any metallic objects. The signature has a different standard deviation for each source-receiver pair acquired by the system. For the second stage, the objective is to discriminate between clutter and UXO while also classifying the type of UXO object. Therefore we require a more accurate representation of the background noise, especially in terms of possible spatial correlation. We included spatially correlated noise by cropping areas from the field data that were labeled as background during the first stage, and then adding these to the synthetic data. To account for a more diverse range of background noise, we applied data augmentation by extracting principal component vectors from randomly cropped samples. We then generated new samples from a multivariate Gaussian distribution, using only the principal components that explained more than 90\% of the variance.

\subsection{Training and validation with synthetic data}
\label{Sec:trainingsynth}

The training of a neural network for classification typically uses the cross-entropy (CE) loss function, which gives a measure of how close are the predicted classes to the truth labels in terms of their probability vectors:

\begin{equation}
\ell_{CE}(p,q) = -\sum_{i=1}^{nc} q_i \log(p_i),
\end{equation}

\noindent where $p_i$ and $q_i$ are the predicted and target probability vectors for class $i$, respectively. In this work, we considered a modified version of this loss function which often performs better for image segmentation when the background class is much more frequent than the other classes. This modified version is termed Focal loss \cite{lin_focal_2018} and it has the form:

\begin{equation}
 \ell_{Focal} = -\sum_{i=1}^{nc} (q_i)^\lambda \log(p_i),
\end{equation}

\noindent where we set a value of $\lambda$ equal to 2 as suggested by \cite{lin_focal_2018}.

The number of training samples and other choices required during training were selected based on a synthetic multi-line dataset. This synthetic multi-line dataset includes three clutter objects that were designed manually with different geometries (i.e. different polarizabilities) and five different UXO objects with different sizes (Figure \ref{Fig:synthbinprobs}). We added synthetic noise that had the same standard deviation signature as observed in the field to our training set. Each source-receiver combination has a different noise signature, and we describe how we accounted for this in the section \ref{Sec:detection} ``Detection and estimation of systematic noise” below. Being a multi-line survey, this example raises some of the challenges that may occur in practice, such as the "halo" effects mentioned above. The following hyperparameters were chosen to reduce classification errors which include both misclassifying objects and incorrectly labeling background as objects (false detections):

\paragraph{Number of training samples} We generated a total of 400,000 synthetic pairs (UltraTEMA-4 data and target labels). Using a smaller number of examples (200,000) decreased the number of points correctly labeled while a higher number (800,000) did not improve multi-class classification results. For the first-phase detection CNN, a lower number of training samples equal to 20,000 was used.

\paragraph{Number of convolutional layers} We performed some tests regarding variations of the base architecture presented in Figure \ref{Fig:architecture}. For example, we tested different numbers of layers in each of the two convolutional stages: 3, 4 and 5. For the synthetic case, we noticed that accuracy in terms of the number of pixels correctly labeled increased when going from 3 to 4 layers while it remained almost constant from 4 to 5 layers. Therefore, we set the number of layers to 4 for the proposed workflow (hence the heading “x 4” in Figure \ref{Fig:architecture}).

\paragraph{Number of UXO classes} We also tested different granularity for the number of classes and we noticed that combining UXOs with similar sizes (for example, 81 mm and medium ISO pipe) increased classification accuracy, at least for the sizes of the training dataset we considered. In this way, we set the number of classes equal to 7 for the multi-class classification.

\paragraph{Distance threshold} The distance threshold in our clutter design strategy was also chosen using this synthetic dataset. According to the histogram of distance shown in Figure \ref{Fig:parspace}a, we considered three different values for distances of 1.0, 1.5 and 2.0. A distance of 2.0 incorrectly classified one of the clutter objects as UXO while the values of 1.0 and 1.5 correctly classified all objects. We choose a value of the distance threshold of 1.5 for our workflow.

\paragraph{Parameters for optimization} The optimization algorithm used was Adam (Kingma and Ba, 2017) which is a modified version of stochastic gradient descent that uses running averages of the gradient; it has been shown to perform better for the training of most deep neural networks. The learning rate (or step length) used was $10^{-3}$ and the parameters to compute the running averages of the gradient were set to their default values ($\beta_1 = 0.9$ and $\beta_2 = 0.999$). The number of epochs (number of iterations where the full training dataset is considered) was set to 20. We also tested using 30 and 40 epochs which increased training time but did not improve performance.

Figure \ref{Fig:synthbinprobs} shows the probability map output for the CNN performing binary classification. These probabilities were obtained by averaging probabilities for all overlapping sliding windows (see subsection \ref{Sec:processingline}). Figure \ref{Fig:synthbinprobs} shows the corresponding classification map (indicating metallic or background). Figure \ref{Fig:synthmultiprobs} shows the probabilities for the multi-class classification, and final classification maps, respectively. We noticed that some areas where no UXOs are located were getting low probabilities of being background but also low probabilities of being either clutter or any type of UXO (e.g. as can be seen by the lower probabilities near the 5m cross-track position in the top panel of Figure 11). To reduce the number of false positives coming from these areas we simply defined a threshold value of 1\% for background classification, that is, any pixel classified as a UXO or clutter but with a probability higher than 1\% of being background was reassigned as background even if its probability of being UXO was higher. Although this seems a small threshold, in practice, we found it to be consistent and very unlikely to cause a UXO to be reassigned as background since the probabilities were always substantially less than 1\% for background when any metallic items were present. The likely reason why these areas are assigned low probabilities of being background is related to the “halo” effects mentioned above (and it was partially solved by considering the strategy of adding objects in the second box in our training dataset) but also to the fact that we are not considering a multi-target scenario for this study. Consider, for example, that we have two objects that are separated by a distance equal to the fixed spatial window (around 4.5 m in our case). In the central part between these two objects we would see the superposition of the signals coming from the two objects. Our CNN was never trained on those instances and it seems to be assigning probability values that are equally distributed throughout all object classes and probabilities for background greater than 1\%. As a result, our current workflow would assign this intermediate region as background. We hypothesize that if we account for multi-target scenarios during training, the impact of this 1\% threshold could be reduced, as the network would have seen these types of scenarios. However, it may be that this remains a useful modification for tackling other types of background effects that may not have been encountered in training.

To solve the issue of false detections for the second CNN, we also tested the possibility of using the first CNN to filter out the regions already classified as background. We tried both multiplying by the probability of the first CNN and applying a background mask. Both of these approaches resulted in slightly cleaner background classification but they gave similar results in terms of the final labels assigned to the objects. The mask approach avoids having to set the 1\% threshold described above, however, it increased the number of false positives by one object for the blind grid case explained below. Considering this, we decided to use the thresholding strategy explained in the previous paragraph to clean the background for the classification part of our workflow.

\begin{figure}
\centering
	\includegraphics[width=0.5\textwidth]{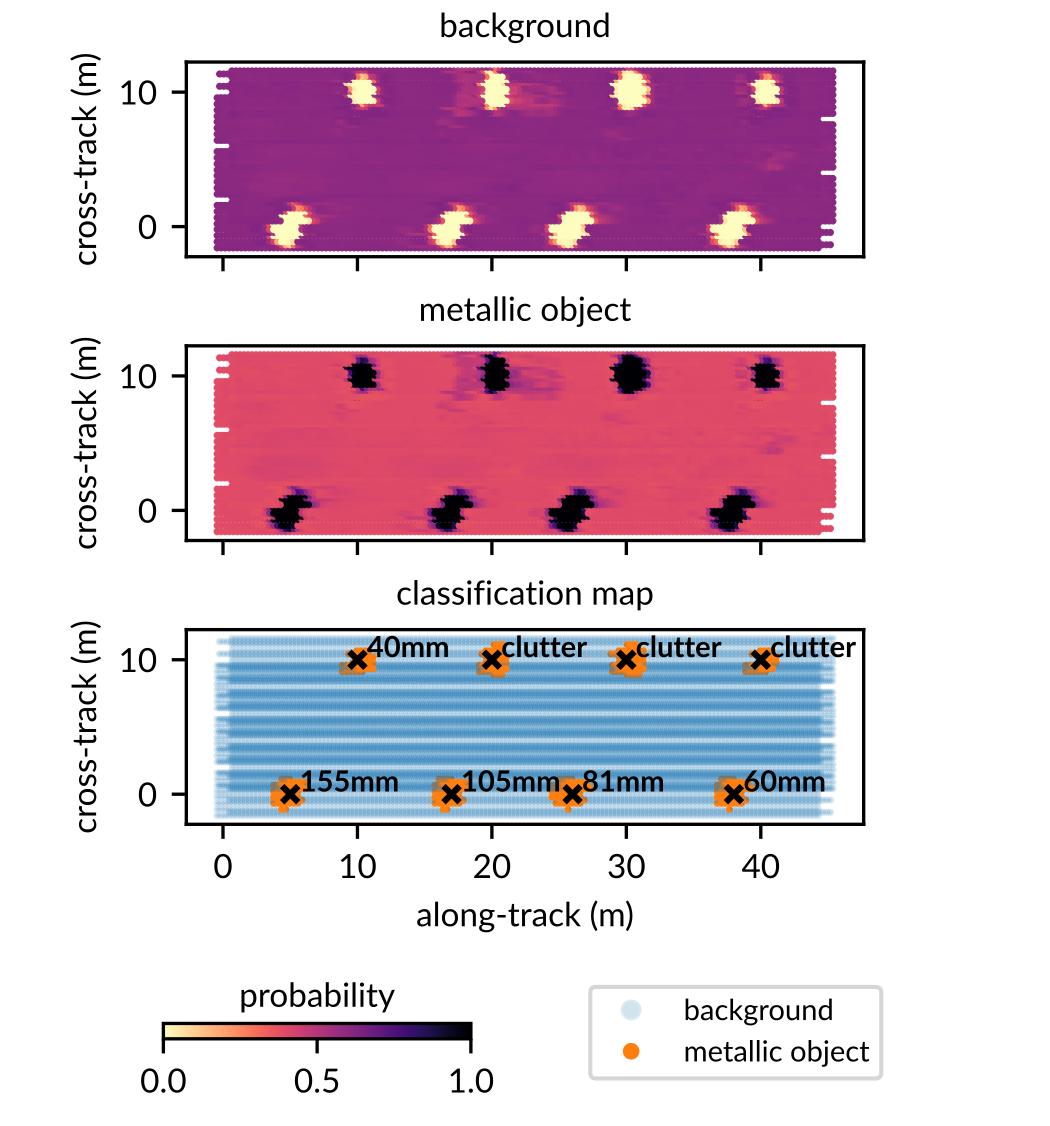}
	\caption{Detection (binary class) probability map outputs the synthetic multi-line dataset (stage 1 of our workflow).}
	\label{Fig:synthbinprobs}
\end{figure}

\begin{figure}
	\centering
	\includegraphics[width=0.5\textwidth]{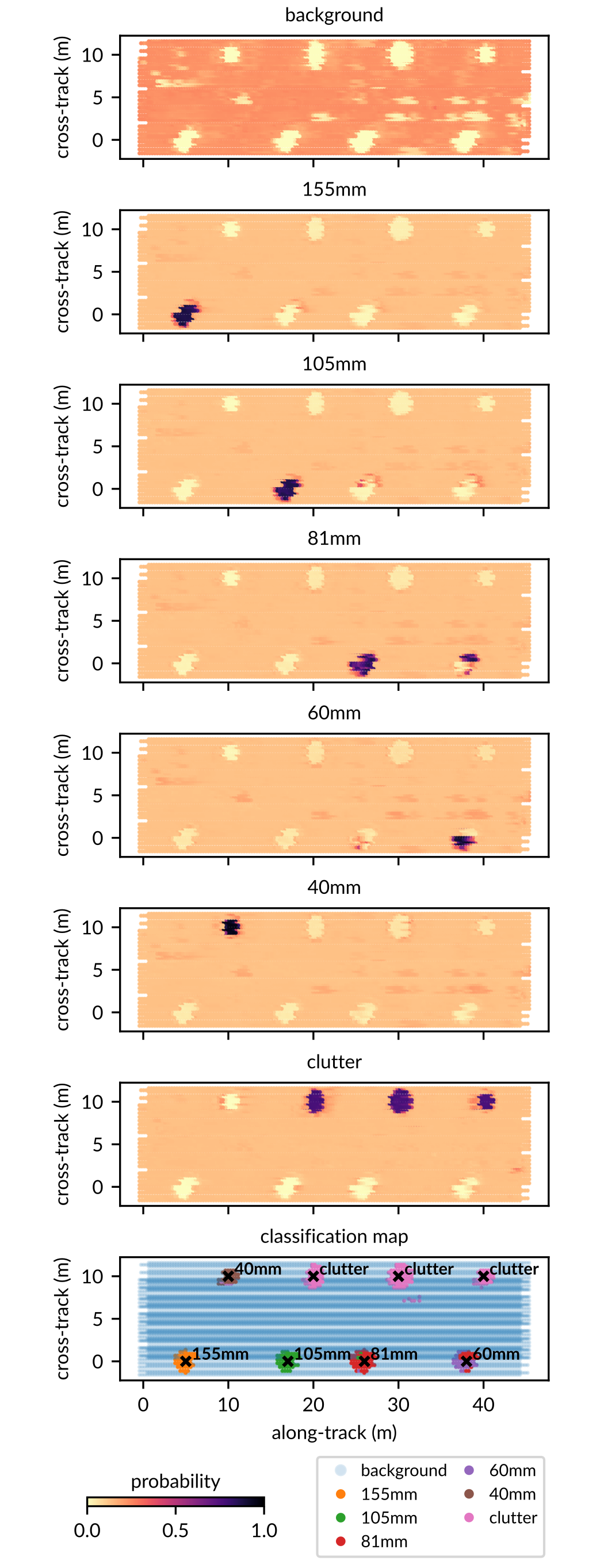}
	\caption{Multi-class probability map outputs the synthetic multi-line dataset (stage 2 of our workflow). The probability of background for the first plot is around 20\% while in the remaining plots, the probability of each class is greater than 90\% in the location of the objects of the corresponding class.}
	\label{Fig:synthmultiprobs}
\end{figure}

\section{Results and Discussion}

\subsection{Classification for calibration line}

\subsubsection{Calibration line data}

We tested our approach using field data acquired with the UltraTEMA-4 system in the Sequim Bay marine test site for years 2021 and 2022 \cite{funk_ultratema-4_2022}. For data acquired in both years, the field dataset includes a calibration grid and a blind grid. Here, we first show data from the calibration grid of 2022 which is composed of 7 acquisition lines and includes 21 seeded objects whose types are known. A preprocessing step was used to remove the EM response of conductive seawater and sediments \cite{song_advanced_2020} but some spatially correlated noise still remains in the preprocessed data. The pre-processed data for the calibration grid of 2022 is shown in Figure \ref{Fig:fielddata}a.

\begin{figure*}
	\centering
	\includegraphics[width=\textwidth]{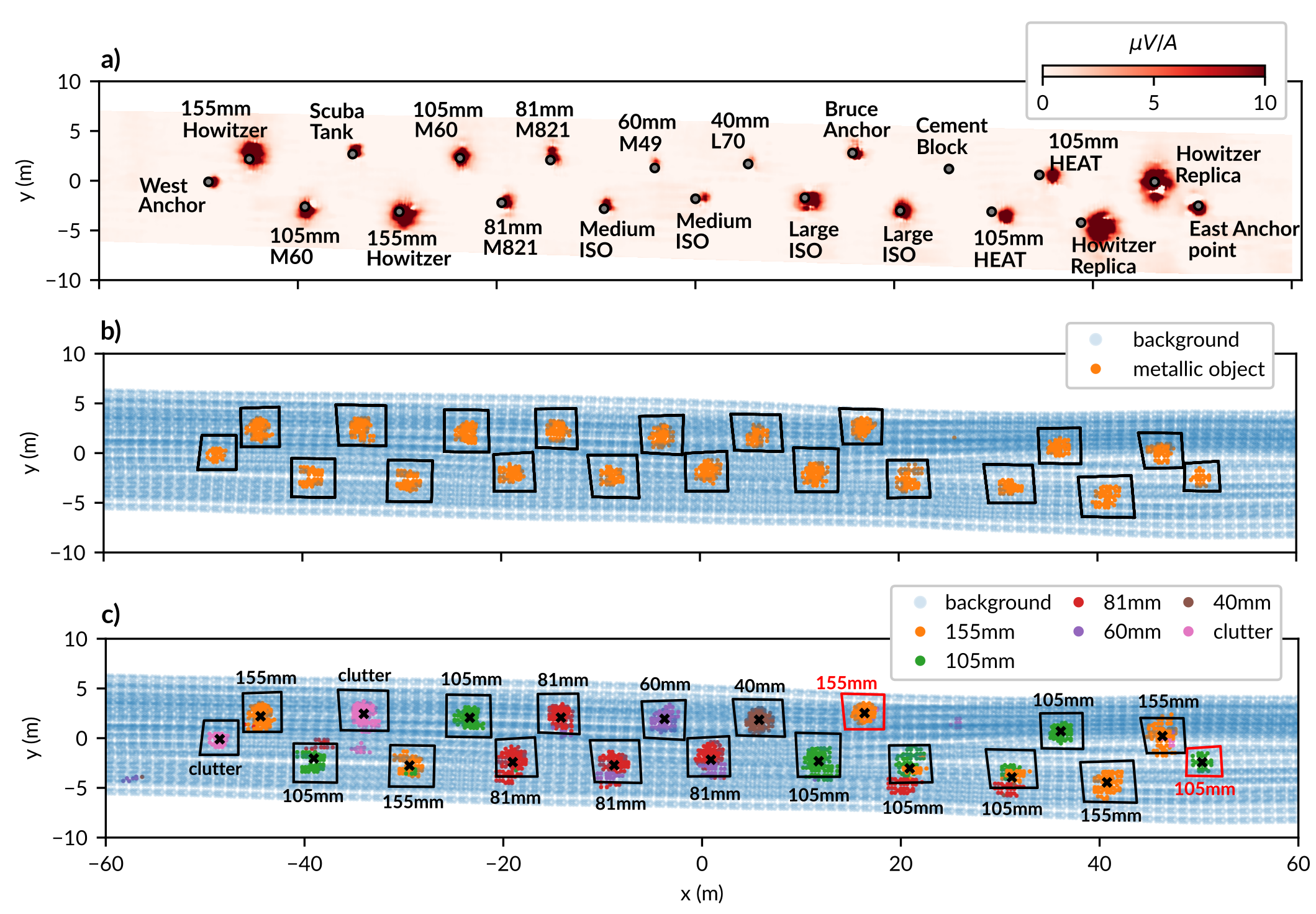}
	\caption{(a) Z-component of EM data (time channel \#5) collected for calibration grid 2022. (b) Detection map (output of first CNN). (c) Classification map (output of second CNN), objects highlighted in red indicate false positives (clutter objects classified as UXOs).}
	\label{Fig:fielddata}
\end{figure*}

\subsubsection{Detection and estimation of systematic noise}
\label{Sec:detection}

As discussed, we adopt a two-stage workflow where we use the first CNN to separate background signal from anomalies (metallic objects) in the field data effectively creating a detection map from which we can pick background signal and estimate spatially correlated (systematic) noise to use during training of our multi-class CNN. For this first detection CNN, we found that it was sufficient to use training samples with the same standard deviation signature of the preprocessed background measurements in the field. This signature is simply the standard deviation given for all time channels and for all transmitter-receiver combinations, which are 144 combinations for the UltraTEMA-4 system (4 transmitters by 12 receivers and 3 components). To get this signature, we manually pick a couple of zones where data seems to be far away from any anomaly (left and right ends of 2022 calibration grid in Figure \ref{Fig:fielddata}a).

Once the detection CNN is trained it produces a classification map from which we can crop out the zones surrounding detected objects and take the remaining data as background signal (Figure \ref{Fig:fielddata}b). Then, we randomly sample pieces of this background signal and estimate the systematic noise to be added to the training dataset of the second CNN.

\subsubsection{Classification results}

Once the multi-class CNN is trained it processes the data per sliding window from which we get a set of probability maps and a single classification map (Figure \ref{Fig:fielddata}c).

While the classification map is already a result from which we can visually identify the class of object, we can further process the information provided by both the probability maps and the classification maps to generate a diglist. The first step in this post-processing is to manually select a cell surrounding each of the classified objects. For each cell, we filter only the output points that are not classified as background and take the corresponding probability values from the probability maps. Then, all selected probabilities are averaged per class within the cell and we choose the class with the highest average probability. The horizontal position is simply estimated by taking the average position in cross- and along-track of the points within the cell that were not classified as background (alternatively, taking a weighted average considering the probability values may improve the estimated position). In this way, we get a single class per object as shown by the labels in Figure \ref{Fig:fielddata}c.

One of the advantages of our proposed workflow is that we can use the probability output of the CNN in post-processing steps. For example, a UXO object in the blind grid analyzed below was labeled as clutter by our CNN but it also had a high probability of being a 105 mm UXO. This means that this object has a high uncertainty for its clutter vs UXO discrimination. To reduce the risk of missing a UXO object, we can use the average probability values to define some rules where we more harshly weight false negatives (classifying UXO as clutter) than false positives (classifying clutter as UXO). For example, we can define a rule where any object that was labeled as clutter but the probability of it being a certain type of UXO is also high, it is re-classified as that UXO class. This value depends on the number of classes. For our field cases, where we considered 7 different classes, we noticed that for a certain object the two classes with highest probability values start to have very similar values when they reach the range of 20-30 percent. To set this safety rule we selected a value of 25\% for all cases, this means that e.g., if an object has a 30\% probability of being clutter and a 25\% probability of being an 81 mm UXO, then we assign the class of 81 mm.

Finally, we noticed that the classes assigned to a small number of objects (less than 10 percent for the field cases considered) changed labels depending on the random initial seed used to train the multi-class CNN. There are two possible reasons for this: (1) modeling of systematic noise is not diverse enough due to a limited number of spatially correlated samples and (2) the gradient-based optimization used for training the CNNs is not guaranteed to reach the same global minimum and some minima may produce slightly more accurate classification results. To fix this, instead of considering the results from a single CNN we considered an ensemble of CNNs, i.e. we trained 5 CNNs initialized with different random seeds and performed a voting scheme on the final class assigned.

\subsection{Classification for blind grid}

Following the same workflow, we also analyze the results for the blind grid data collected in 2021. This dataset is composed of 17 acquisition lines and includes 35 seeded objects whose types are known (18 clutter objects and 17 UXOS). The preprocessed data with ground truth labels is shown in Figure \ref{Fig:blind2021}a while the classification results are shown in Figure \ref{Fig:blind2021}b. For this dataset, UXO objects are always separated by a minimum distance of 5 m, therefore we can consider that it is not a multi-target scenario. The classification map shows 5 clutter objects classified as UXOs (false positives). We noted that the safety rule explained above helped fix only one instance where a UXO was classified as clutter (the rest of the UXO were correctly classified as UXOs). Note that some clutter objects were classified as a single one when they were too close together and some clutter objects were not detected. Neither of those scenarios indicates a failure of the classifier since we are interested in (1) detecting and correctly classifying all UXOs (maximizing true positives), and (2) among the detected objects, classifying most of the clutter objects as such (minimizing false positives).

Comparing these results with those of a physics-based approach, both methods correctly detected and classified all UXOs. Figure \ref{Fig:ROC} shows the ROC (Receiver Operating Characteristic) curve for both approaches. When comparing the best threshold (point in the ROC curve where all UXOs have been dug), both approaches reach a value of 5 false positives. This shows that the two approaches have similar performance for the single target scenario. Note that data quality has a similar impact in both approaches: both require background removal first and the remaining noise may be handled with both approaches without diminishing accuracy.

\begin{figure*}
	\centering
	\includegraphics[width=\textwidth]{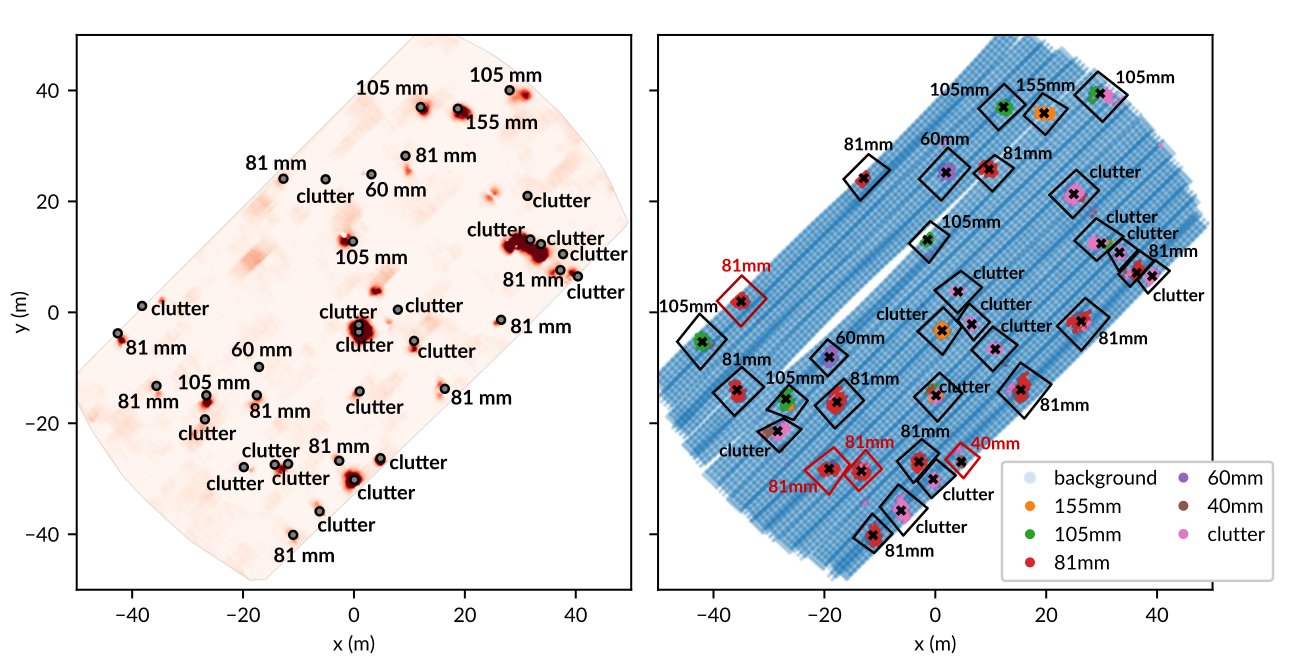}
	\caption{(a) Z-component of EM data (time channel \#5) collected for blind grid 2021. (b) Classification map (output of second CNN), objects highlighted in red indicate false positives (clutter objects classified as UXOs).}
	\label{Fig:blind2021}
\end{figure*}

\begin{figure}
	\centering
	\includegraphics[width=0.5\textwidth]{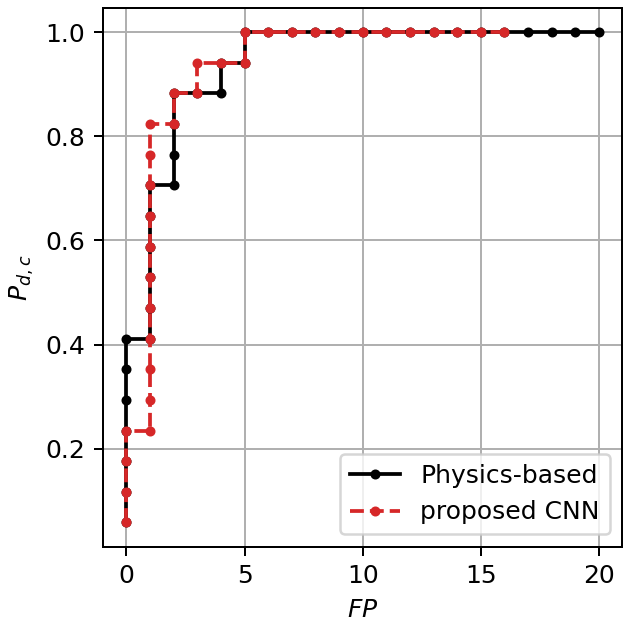}
	\caption{ROC curves for blind grid 2021. $FP$ indicates the number of false positives and $P_{d,c}$ indicates the probability of detection and classification of all the UXOs.}
	\label{Fig:ROC}
\end{figure}

\subsection{Discussion}

The workflow developed in this study is the first that we are aware of to apply neural networks directly to EMI data for UXO classification. Although still at an experimental stage, the CNN-based workflow achieves classification accuracy approaching that of the standard inversion-based approach, demonstrating its potential as a complementary classification tool. Once trained, the CNN enables rapid classification with minimal human supervision. Agreement between the CNN and inversion-based results can increase confidence in a classification, whereas disagreement can identify candidates that warrant closer examination. 

A key feature of our CNN is its segmentation-like architecture, which provides high-resolution classification maps that do not require interpolation and are therefore less likely to miss small objects. This approach does require an appropriate way to define the output targets, i.e. the label masks. We found that the mask, defined by setting a threshold on the normalized synthetic EM data, needs to be large enough that the smallest objects considered have at least 20 pixels different from the background, for the cases considered in this study.

The clutter design is another important element of our approach. As with any machine learning method, the CNN must be trained with examples of the UXO expected at the site as well as the range of possible clutter objects. The physics-based parameterization used here (Equation \eqref{Eq:pasion}) aims to cover a broad range of clutter objects by setting a distance threshold in the parameter space. This distance must be tuned to maximize clutter discrimination without reducing UXO detection, and it may be chosen using a line of synthetic data that includes both UXO and objects with the expected dimensions of clutter and/or a field calibration line of the kind typically available for UXO clearance projects.

Several other choices improved the performance of the classification CNN, including grouping UXO of similar size into a single class and averaging probabilities over overlapping sliding windows. The addition of spatially correlated noise, discussed above, was the most significant of these. Note that our strategy to get this systematic noise means that a CNN will need to be trained for each new set of collected data, unless it is for the same site and with the same instrument. For most of the sites, a calibration line is a standard practice from which some initial estimate of background noise is feasible. Future work will address this limitation by e.g. forward modeling background response knowing the position and attitude of the EMI system and the structure of shallow marine sediments and their electrical resistivity. The structure may include spatially correlated random variations to simulate real-world variability. Afterwards, we will apply the same background removal techniques and we expect to end with spatially correlated noise similar to the one seen on the field. Although there might be other sources of systematic noise, we expect them to have a lower magnitude and hence have a minor impact on the results.

One area that requires future work is multi-target scenarios, in which two or more objects fall within the same spatial window. Results for another dataset (blind grid 2022; not shown) that included several such instances showed reduced classification accuracy compared to the inversion-based approach. This scenario has already been addressed in the traditional approach using a variety of strategies \cite{hu_classification_2004,song_nonlinear_2011,song_estimating_2012,t_m_grzegorczyk_simultaneous_2011,grzegorczyk_subsurface_2012}. Outside of multi-target scenarios, the accuracy of our approach approaches that of the inversion-based workflow.

We also see opportunities for improving accuracy through the design of the training set and the procedure used to train the neural network. Classification accuracy may be increased by improving the estimation of spatially correlated noise (e.g., by augmenting its diversity using a forward model for stratified media with variable orientation of the system) and possibly by using other regularization methods for CNNs such as dropout or L2 regularization. There are also opportunities to automate some steps of our workflow that are currently manual, such as picking the objects from the detection map in order to then estimate background noise; this should be a relatively straightforward problem to address with existing algorithms. Training may be completed before the main survey when a representative calibration line or sufficient information about background noise is available, thereby enabling the trained CNN to support rapid first-pass or near-real-time classification in conjunction with the standard inversion-based workflow.

\section{Conclusions}

Modern EMI systems used for UXO classification collect spatially dense data with sufficient information to classify different types of UXO and discriminate clutter. The traditional approach for classification is based on inversion and polarizability curve matching. We developed a CNN-based approach for UXO classification directly from EMI data. Our workflow uses two CNNs, the first one for anomaly detection and the second one for classification. The architectures follow an image segmentation structure where the output preserves spatial dimensions of the input and therefore produces high resolution classification maps. The CNNs were trained with synthetic data using a dipole model and considering UXOs and clutter objects. We proposed a new strategy to design clutter in a physics-based parameter space allowing to cover a broad range of objects which is crucial for machine learning methods. Spatially and temporally correlated noise, that is consistent with that of the field, is added to the training dataset to improve classification results. Our workflow was applied to marine EMI datasets from a test site where it successfully detected all of the UXOs while discriminating most of the clutter. The CNN has not been trained for multi-target scenarios (two or more objects within the same spatial window) hence it leads to errors for such cases. In future work, we plan to explore this multi-target scenario by including these cases in the training dataset of our CNN. Moreover, we will further test its performance and generalizability into new real field conditions, and explore the sensitivity to the noise characteristics provided during training. To the best of our knowledge, this work is the first example of using neural networks to classify UXO directly from multicomponent EMI data. Having an end-to-end workflow will enable us to assess the strengths of a CNN-based approach and identify opportunities where components of our workflow may complement or offer improvements to the standard physics-based workflow.

\section*{Acknowledgements}

This work was supported by the Strategic Environmental Research and Development Program (SERDP) under the SERDP Exploratory Development (SEED) project MR22-3487.

\section*{Data availability}
The modelling and field data supporting the findings of this study are available
from the corresponding author upon reasonable request.

\section*{Code availability}

The code implementation of the full workflow is available on \href{https://github.com/jlalvis/uxo-cnn-classification}{https://github.com/jlalvis/uxo-cnn-classification}.

\bibliographystyle{apalike}
\bibliography{UXO_project}

\end{document}